\documentclass[10pt,letterpaper,journal,twocolumn]{IEEEtran}

\usepackage{amssymb}
\usepackage[noadjust]{cite}

\newtheorem{IEEE-p}{Proposition}[section]
\newtheorem{IEEE-t}[IEEE-p]{Theorem}
\newtheorem{IEEE-d}[IEEE-p]{Definition}
\newtheorem{IEEE-c}[IEEE-p]{Corollary}
\newtheorem{IEEE-l}[IEEE-p]{Lemma}
\newtheorem{IEEE-r}[IEEE-p]{Remark}
\newtheorem{IEEE-e}[IEEE-p]{Example}

\newcommand\myproposition[1]{\smallskip\begin{IEEE-p}#1\end{IEEE-p}}
\newcommand\mytheorem[1]{\smallskip\begin{IEEE-t}#1\end{IEEE-t}}
\newcommand\mydefinition[1]{\smallskip\begin{IEEE-d}#1\end{IEEE-d}}
\newcommand\mycorollary[1]{\smallskip\begin{IEEE-c}#1\end{IEEE-c}}
\newcommand\mylemma[1]{\smallskip\begin{IEEE-l}#1\end{IEEE-l}}
\newcommand\myremark[1]{\smallskip\begin{IEEE-r}#1\end{IEEE-r}}
\newcommand\myexample[1]{\smallskip\begin{IEEE-e}#1\end{IEEE-e}}
\renewcommand{\QED}{\QEDopen}
\newcommand\myproof[1]{\noindent\hspace{2em}{\it Proof: } #1 \endproof}

\renewcommand{\leq}{\leqslant}
\renewcommand{\geq}{\geqslant}

\def\mod{{\mathrm{\phantom{|}mod\phantom{i}}}}

\def\N{{\mathbb {N}}}
\def\F{{\mathbb {F}}}
\def\Fq{{\mathbb {F}}_q}
\def\mF{{\mathcal {F}}}
\def\mL{{\mathcal {L}}}

\newcommand{\casos}[4]{
\left\{
\begin{array}{ll}
#1 & \mbox{ #2 }\\
#3 & \mbox{ #4 }\\
\end{array}
\right.
}
\newcommand{\trescasos}[6]{
\left\{
\begin{array}{ll}
#1 & \mbox{ #2 }\\
#3 & \mbox{ #4 }\\
#5 & \mbox{ #6 }\\
\end{array}
\right.
}
\newcommand{\quatrecasos}[8]{
\left\{
\begin{array}{ll}
#1 & \mbox{ #2 }\\
#3 & \mbox{ #4 }\\
#5 & \mbox{ #6 }\\
#7 & \mbox{ #8 }\\
\end{array}
\right.
}

\newcommand{\black}[1]{#1}

\begin{document}

\date{\today}
\title{Acute Semigroups,
the Order Bound on the Minimum Distance
and the Feng-Rao Improvements}

\author{Maria Bras-Amor\'{o}s$^*$
\thanks{
This work was supported in part by the Spanish CICYT 
under Grant TIC2003-08604-C04-01, 
by Catalan DURSI under Grant 2001SGR 00219.
\newline\indent
The author is with the Computer Science Department, Universitat
Aut\`onoma de Barcelona, 08193-Bellaterra, Catalonia, Spain (e-mail: mbras@ccd.uab.es).
}
}

\markboth{Manuscript accepted in IEEE Transactions On Information Theory, Published in Vol. 50, No. 6, June 2004}
{Maria Bras-Amor\'os: Acute Semigroups, 
the Order Bound on the Min. Dist.
and the Feng-Rao Improvements}

\maketitle

\begin{abstract}
We introduce
a new class of numerical semigroups,
which we call the class of
{\it acute} semigroups
and we prove that they generalize
symmetric and pseudo-symmetric numerical semigroups,
Arf numerical semigroups
and the semigroups generated by an interval.
For a numerical semigroup $\Lambda=\{\lambda_0<\lambda_1<\dots\}$
denote $\nu_i=\#\{j\mid\lambda_i-\lambda_j\in\Lambda\}$.
Given an acute numerical semigroup $\Lambda$
we find the smallest non-negative integer $m$
for which
the order bound on the minimum distance
of one-point Goppa codes
with associated semigroup $\Lambda$
satisfies
$d_{ORD}(C_i)(:=\min\{\nu_j\mid j>i\})=\nu_{i+1}$
for all $i\geq m$.
We prove that the only numerical semigroups for which
the sequence $(\nu_i)$ is always non-decreasing
are ordinary numerical semigroups.
Furthermore we show that a semigroup can be uniquely determined by
its sequence $(\nu_i)$.
\end{abstract}


\section{Introduction}
Let $\N_0$ denote the set of all non-negative integers.
A {\it numerical semigroup} is a subset $\Lambda$
of $\N_0$
containing $0$, closed
under summation and with finite complement in $\N_0$.
For a numerical semigroup $\Lambda$
define
the {\it genus} of $\Lambda$ as the number
$g=\#(\N_0\setminus\Lambda)$
and the {\it conductor} of $\Lambda$ as
the unique integer
$c\in\Lambda$
such that $c-1\not\in\Lambda$ and $c+\N_0\subseteq\Lambda$.
The elements in $\Lambda$ are called the {\it non-gaps} of $\Lambda$
while the elements in
$\Lambda^c=\N_0\setminus\Lambda$ are called the {\it gaps} of $\Lambda$.
The {\it enumeration}
of $\Lambda$ is the unique increasing bijective map
$\lambda:\N_0\longrightarrow\Lambda$.
We will use $\lambda_i$ for $\lambda(i)$.
For further details on numerical semigroups we refer the reader to \cite{HoLiPe:agc}.
The first aim of this work is to give a new class of
numerical semigroups containing the
well-known classes of symmetric and pseudo-symmetric semigroups,
Arf semigroups and
semigroups generated by an interval and
which in turn is not the whole set of numerical semigroups.

Let $F/\F$ be a function field and let $P$ be
a rational point of $F/\F$.
For a divisor $D$ of $F/\F$,
let $\mL(D)=\{0\}\cup\{f\in F^*\mid (f)+D\geq 0\}$.
Define $A=\bigcup_{m\geq 0}\mL(m P)$ and
let $\Lambda=\{-v_P(f)\mid f\in A\setminus\{0\}\}=\{-v_i\mid i\in\N_0\}$ with
$-v_i<-v_{i+1}$.
It is well known that the number of elements in $\N_0$
which are not in $\Lambda$ is equal to the genus of
the function field.
Moreover,
$v_P(1)=0$ and $v_P(fg)=v_P(f)+v_P(g)$
for all $f,g\in A$.
Hence, $\Lambda$
is a numerical semigroup. It is called the
{\it Weierstrass semigroup} at $P$.

Suppose $P_1,\dots,P_n$ are
pairwise distinct rational points of $F/\Fq$ which are different from $P$
and let $\varphi$ be the map
$A\rightarrow\Fq^n$ such that $f\mapsto(f(P_1),\dots,f(P_n))$.
For $m\geq 0$ the
{\it one-point Goppa code}
of order $m$
associated to $P$ and $P_1,\dots,P_n$
is defined as
$C_m=\varphi(\mL(\lambda_mP))^\perp$.

Suppose that $\lambda$
is the enumeration of $\Lambda$.
For any $i\in\N_0$
let $N_i=\{j\in\N_0\mid \lambda_i-\lambda_j\in\Lambda\}$
and let $\nu_i=\# N_i$.
The {\it order bound} on the minimum distance
of the code $C_m$
is defined as
$$d_{ORD}(C_m)=\min\{\nu_i\mid i>m\}$$
and it satisfies $d_{C_m}\geq d_{ORD}(C_m)$
where $d_{C_m}$ is the minimum distance
of the code $C_m$.
A better bound is the {\it refined order bound}
defined as
$$d_{ORD}^\varphi(C_m)=\min\{\nu_i\mid i>m, C_i\neq C_{i-1}\}.$$
This one satisfies
$d_{C_m}\geq d_{ORD}^\varphi(C_m)\geq d_{ORD}(C_m)$.
The order bound depends only on the 
Weierstrass semigroup whereas the refined order bound 
depends on the Weierstrass semigroup and on the
map $\varphi$ as well.
These bounds can be found in 
\cite{FeRa:dFR,KiPe:telescopic,HoLiPe:agc}.
We have that for all $i\geq\lambda^{-1}(2c-1)$,
the sequence $(\nu_i)$ is increasing and so,
for all $i\geq\lambda^{-1}(2c-1)-1$,
$d_{ORD}(C_i)=\nu_{i+1}$.
This equality makes it easier to
compute the order bound, since we do not need to find
a minimum in a set.
The second goal of this work is to find
an explicit formula for
the smallest $m$ for which
$d_{ORD}(C_i)=\nu_{i+1}$ for all
$i\geq m$.
This is equivalent to search for
the largest $i$ at which $(\nu_i)$ is decreasing.
We will give such a formula
for the
numerical
semigroups in the class of {\it acute} semigroups.

On the other hand,
given a designed minimum distance $\delta\in\N_0$,
since $d_{ORD}(C_m)=\min\{\nu_i\mid i>m\}$,
the code $C_m$ with largest dimension
among the codes $C_i$
with $d_{ORD}(C_i)\geq \delta$
is with $m=m(\delta)=\max\{i\in\N_0\mid \nu_i<\delta\}.$
We call this code $C(\delta)$.
Now, let $\mF=\{f_i\in A\mid i\in\N_0\}$
be such that $v_P(f_i)=v_i$.
With this notation,
$$C_m=[\varphi(f_i)\mid i\leq m]^\perp,$$
where $[u_1,\dots,u_n]$ is the $\F_q$-vector space 
spanned by $u_1,\dots,u_n$.
Define
$$\widetilde{C}(\delta)=[\varphi(f_i)\mid \nu_i<\delta]^\perp.$$
This is a code and it satisfies $\dim\widetilde{C}(\delta)\geq\dim C(\delta)$
because
$\{i\in\N_0\mid\nu_i<\delta\}\subseteq\{i\in\N_0\mid i\leq m(\delta)\}$.
Feng and Rao proved that the minimum distance of
$\widetilde{C}(\delta)$
is larger than or equal to $\delta$
and hence, they are an improvement
to one-point Goppa codes
\cite{FeRa:improved}.
If moreover we take the morphism $\varphi$
into account we can drop the redundant rows in the parity check martix
and define the code
$\widetilde{C}_\varphi(\delta)=[\varphi(f_i)\mid \nu_i<\delta,C_i\neq C_{i-1}]^\perp$.
The third goal of this work is to
characterize the
numerical semigroups
that satisfy
$$\{i\in\N_0\mid\nu_i<\delta\}\subsetneq\{i\in\N_0\mid i\leq m(\delta)\}$$
at least for one value of $\delta$.
Notice that this is equivalent to ask for
which numerical semigroups
the sequence $(\nu_i)$ 
has $\nu_i>\nu_{i+1}$ for at least one $i$.

This will give a characterization of
a class of semigroups
by means of a property on the sequence $(\nu_i)$.
We may ask how a numerical semigroup can be
determined by its associated sequence
$(\nu_i)$. The last goal of this work
is to prove that any numerical semigroup is
uniquely determined by its sequence $(\nu_i)$.

In Section~\ref{section: symmetric numerical semigroups}
we give the definition of symmetric and pseudo-symmetric 
semigroup and some related
known results \cite{KiPe:telescopic,HoLiPe:agc,RoBr:irreducible},
in Section~\ref{section: Arf numerical semigroups}
we define Arf numerical semigroups as in \cite{CaFaMu:arf}
and give some results and in
Section~\ref{section: interval numerical semigroups}
we present numerical semigroups generated by an interval as
in \cite{GaRo:interval}.
In Section~\ref{section: acute numerical semigroups}
we introduce the definition of
acute numerical semigroups and we prove that they include
symmetric, pseudo-symmetric, Arf and interval-generated semigroups.
We then find in Section~\ref{section: frd acute}, for
acute numerical semigroups,
the largest $m\in\N_0$
for which $\nu_m>\nu_{m+1}$
and hence the smallest $m$ for which $d_{ORD}(C_i)=\nu_{i+1}$
for all $i\geq m$.
We prove in Section~\ref{section: nu no decreasing implica ordinary}
that the only numerical semigroups for which $(\nu_i)$ is always non-decreasing,
that is, $d_{ORD}(C_i)=\nu_{i+1}$ for all $i\in\N_0$,
or equivalently
$\{i\in\N_0\mid \nu_i<\delta\}=\{i\in\N_0\mid i\leq m(\delta)\}$
for all $\delta\in\N_0$,
are
ordinary numerical semigroups.
Those are the numerical semigroups that are equal to
$\{0\}\cup\{i\in\N_0\mid i\geq c\}$ for some non-negative integer $c$.
We finally prove in Section~\ref{section: nu determina lambda} that
a numerical semigroup can be uniquely determined by its sequence $(\nu_i)$.

\section{Symmetric and pseudo-symmetric numerical semigroups}
\label{section: symmetric numerical semigroups}

\mydefinition{A numerical semigroup $\Lambda$ with genus $g$ and conductor $c$ is said to be
{\it symmetric} if $c=2g$.
\index{semigroup!symmetric}}

Symmetric numerical semigroups have been studied in
\cite{KiPe:telescopic,HoLiPe:agc,CaFa:semigroup_singular_plane_models}.

\myexample{
\label{exemple: semigroups gen by two int}
Semigroups
{\it generated by two integers}
\index{semigroup!generated by two integers}
are the semigroups of the form
$$\Lambda=\{ma+nb\mid a,b\in\N_0\}$$
for some integers $a$ and $b$.
For $\Lambda$ having 
finite complement in $\N_0$
it is necessary that $a$ and $b$ are
coprime integers.
Semigroups
generated by two coprime integers are 
symmetric \cite{KiPe:telescopic,HoLiPe:agc}.

Geil introduces in
\cite{Geil:Norm-trace-codes}
the norm-trace curve over ${\mathbb F}_{q^r}$
defined by the affine equation
$$x^{(q^r-1)/(q-1)}
=y^{q^{r-1}}+y^{q^{r-2}}
+\dots+y$$
where $q$ is a prime power.
It has a single rational point at infinity
and the Weierstrass semigroup at
the rational point at infinity
is generated by the two coprime integers
$(q^r-1)/(q-1)$ and
$q^{r-1}$.
So, it is an example
of a symmetric numerical semigroup.

Properties on semigroups
generated by two coprime integers can
be found in \cite{KiPe:telescopic}.
For instance, the
semigroup generated by $a$ and $b$,
has conductor equal to $(a-1)(b-1)$,
and any element $l\in\Lambda$
can be written uniquely as
$l=ma+nb$ with $m,n$
integers such that $0\leq m<b$.

From the results in \cite[Section 3.2]{HoLiPe:agc}
one can get, for any
numerical semigroup $\Lambda$ generated by two integers,
the equation of a curve having a point whose
Weierstrass semigroup is $\Lambda$.
}

Let us state now a proposition related
to symmetric numerical semigroups.

\myproposition{
\label{proposition:symmetric-caracteritzacio}
\label{proposition:symmetric-implicacio}
A numerical semigroup $\Lambda$
with conductor $c$
is symmetric if and only if
for any non-negative integer $i$, if $i$ is a gap,
then $c-1-i$ is a non-gap.
}

The proof
can be found in
\cite[Remark~4.2]{KiPe:telescopic}
and \cite[Proposition~5.7]{HoLiPe:agc}.
It follows by counting the number of gaps and non-gaps
smaller than the conductor and the fact that 
if $i$ is a non-gap then $c-1-i$ must be a 
gap because otherwise $c-1$ would also be a non-gap.

\mydefinition{A numerical semigroup $\Lambda$ with genus $g$ and conductor $c$
is said to be {\it pseudo-symmetric}
if $c=2g-1$.}

Notice that a symmetric numerical semigroup can not be pseudo-symmetric.
Next proposition as well as its proof is analogous to 
Proposition~\ref{proposition:symmetric-implicacio}. 

\myproposition{
A numerical semigroup $\Lambda$
with odd conductor $c$
is pseudo-symmetric if and only if
for any non-negative integer $i$ different from $(c-1)/2$, if $i$ is a gap,
then $c-1-i$ is a non-gap.
}

\myexample{
\label{example:klein}
The Klein quartic
\index{Klein quartic!curve}
over $\F_q$
is defined by the affine equation
$$x^3y+y^3+x=0$$
and it is non-singular
if $\gcd(q,7)=1$.
Suppose $\gcd(q,7)=1$ and
denote $P_0$ the
rational point with
affine coordinates $x=0$ and $y=0$.
The Weierstrass semigroup
\index{Klein quartic!semigroup}
at $P_0$
is
$$\Lambda=\{0,3\}\cup\{i\in\N_0\mid i\geq 5\}.$$
For these results we refer the reader to
\cite{Pretzel,HoLiPe:agc}.
In this case $c=5$
and the only gaps different from $(c-1)/2$ are
$l=1$ and $l=4$. In both cases we have
$c-1-l\in\Lambda$. This proves that
$\Lambda$ is pseudo-symmetric. 
}

In \cite{RoBr:irreducible} the authors prove that
the set of irreducible semigroups,
that is, the semigroups that can not be expressed as a proper intersection
of two numerical semigroups,
is the union of the set of symmetric semigroups and 
the set of pseudo-symmetric semigroups.

\section{Arf numerical semigroups}
\label{section: Arf numerical semigroups}

\mydefinition{A numerical semigroup $\Lambda$
with enumeration $\lambda$
is called an {\it Arf numerical semigroup}
\index{semigroup!Arf}
if
$\lambda_i+\lambda_j-\lambda_k\in\Lambda$
for every $i,j,k\in \N_0$ with $i\geq j\geq k$
\cite{CaFaMu:arf}.}

For further work on Arf numerical semigroups we refer the reader to
\cite{BaDoFo,RoGaGaBr}.
For results on Arf semigroups related to
coding theory, see \cite{Bras:AAECC,CaFaMu:arf}.

\myexample{
It is easy to check that the Weierstrass semigroup
in Example~\ref{example:klein} is Arf.
}

Let us state now two results
on Arf numerical semigroups
that will be used later.

\mylemma{
\label{lemma: arf te els non-gaps separats}
Suppose $\Lambda$ is Arf. If $i,i+j\in\Lambda$
for some $i,j\in\N_0$,
then
$i+kj\in\Lambda$ for all $k\in\N_0$.
Consequently,
if $\Lambda$ is Arf and $i,i+1\in\Lambda$, then $i\geq c$.}

\myproof{Let us prove this by induction on $k$.
For $k=0$ and $k=1$ it is obvious.
If $k>0$ and $i,i+j,i+kj\in\Lambda$ then $(i+j)+(i+kj)-i=i+(k+1)j\in\Lambda$.}

Let us give the definition
of inductive numerical semigroups.
They are an example of Arf numerical semigroups.

\mydefinition{A sequence $(H_n)$ of
numerical semigroups is called {\it inductive}
if there exist sequences $(a_n)$ and $(b_n)$
of positive integers such that $H_1=\N_0$ and for
$n>1$, $H_n=a_nH_{n-1}\cup\{m\in\N_0\mid m\geq a_nb_{n-1}\}$.
A numerical semigroup is called {\it inductive} if it is a member of an
inductive sequence
\cite[Definition 2.13]{PeTo:redundancy_improved_codes}.
\index{semigroup!inductive}}

\myproposition{Inductive numerical semigroups are Arf.}
\myproof{
\cite{CaFaMu:arf}.
}

\myexample{
\label{example:2nd GS es Arf}
Pellikaan, Stichtenoth and Torres proved in \cite{PeStTo}
that the
numerical
semigroups
for the
codes
over ${\mathbb F}_{q^2}$
associated to the second tower of Garcia-Stichtenoth
\index{Garcia-Stichtenoth towers!semigroup}
attaining the Drinfeld-Vl\u adu\c t
bound
\cite{GaSt:tff}
are given
recursively by $\Lambda_1=\N_0$
and, for $m>0$,
$$\Lambda_{m}=q\cdot\Lambda_{m-1}\cup\{i\in\N_0\mid i\geq
q^m-q^{\lfloor(m+1)/2\rfloor}\}.$$
They
are examples of inductive numerical semigroups
and hence, examples of Arf numerical semigroups.
}

\myexample{
{\it Hyperelliptic numerical semigroups}.
\index{semigroup!hyperelliptic}
These are the
numerical semigroups
generated by $2$ and an odd integer.
They are of the form $$\Lambda=\{0,2,4,\dots,2k-2,2k,2k+1,2k+2,2k+3,\dots\}$$
for some positive integer $k$.}

\myproposition{The only Arf symmetric semigroups are hyperelliptic semigroups.}
\myproof{\cite[Proposition~2]{CaFaMu:arf}.}

In order to show which are the only Arf pseudo-symmetric semigroups
we need the following definition and lemma.

\mydefinition{
Let $\Lambda$ be a numerical semigroup. The Ap\'ery set of $\Lambda$ is
$$Ap(\Lambda)=\{l\in\Lambda\mid l-\lambda_1\not\in\Lambda\}.$$
}

\myremark{
\label{remark:Ap}
$\# Ap(\Lambda)=\lambda_1$.
}

\mylemma{
\label{lemma:pseudo}
Let $\Lambda$ be a pseudo-symmetric numerical semigroup. For any
$l\in Ap(\Lambda)$ 
different from $\lambda_1+(c-1)/2$,
$\lambda_1+c-1-l\in Ap(\Lambda)$.
}

\myproof{
Let us prove first that 
$\lambda_1+c-1-l\in \Lambda$.
Since 
$l\in Ap(\Lambda)$,
$l-\lambda_1\not\in\Lambda$ and it is different from 
$(c-1)/2$ by hypothesis. Thus
$\lambda_1+c-1-l=c-1-(l-\lambda_1)\in \Lambda$ 
because $\Lambda$ is pseudo-symmetric. 

Now, $\lambda_1+c-1-l-\lambda_1=c-1-l\not\in\Lambda$ 
because otherwise $c-1\in\Lambda$. So
$\lambda_1+c-1-l$ must belong to $Ap(\Lambda)$.
}

\myproposition{
The only Arf pseudo-symmetric semigroups are
$\{0,3,4,5,6,\dots\}$ and
$\{0,3,5,6,7,\dots\}$ (corresponding to the Klein quartic).
}

\myproof{
Let $\Lambda$ be an Arf pseudo-symmetric
numerical semigroup.
Let us show first that 
$Ap(\Lambda)=\{0,\lambda_1+(c-1)/2,\lambda_1+c-1\}$.
The inclusion $\supseteq$ is obvious. In order to prove the opposite inclusion 
suppose $l\in Ap(\Lambda)$, $l\not\in\{0,\lambda_1+(c-1)/2,\lambda_1+c-1\}$.
By Lemma~\ref{lemma:pseudo}, $\lambda_1+c-1-l\in \Lambda$
and since $l\neq\lambda_1+c-1$, 
$\lambda_1+c-1-l\geq\lambda_1$.
On the other hand,
if $l\neq 0$ then $l\geq \lambda_1$.
Now, by the Arf condition, 
$\lambda_1+c-1-l+l-\lambda_1=c-1\in\Lambda$, which is a contradiction.

Now, if $\# Ap(\Lambda)=1$ then, by Remark~\ref{remark:Ap}, $\lambda_1=1$
and $\Lambda=\N_0$. But $\N_0$ is not pseudo-symmetric.

If $\# Ap(\Lambda)=2$ then, by Remark~\ref{remark:Ap}, $\lambda_1=2$.
But then $\Lambda$ must be hyperelliptic and so $\Lambda$ is not 
pseudo-symmetric.

So $\# Ap(\Lambda)$ must be $3$.
Now Remark~\ref{remark:Ap} implies that $\lambda_1=3$ and that 
$1$ and $2$ are gaps.
If $1=(c-1)/2$ then $c=3$ and this gives $\Lambda=\{0,3,4,5,6,\dots\}$.
Else if $2=(c-1)/2$ then $c=5$ and this 
gives $\Lambda=\{0,3,5,6,7,\dots\}$.
Finally, if 
$1\neq(c-1)/2$ 
and 
$2\neq(c-1)/2$,
since $\Lambda$ is pseudo-symmetric,
$c-2,c-3\in\Lambda$. 
But this contradicts 
Lemma~\ref{lemma: arf te els non-gaps separats}.
}

The next two propositions are two characterizations of
Arf numerical semigroups.

\myproposition{
\label{proposition: charactarf 2j-i}
The numerical semigroup $\Lambda$ with
enumeration $\lambda$ is Arf if and only if for every two positive integers $i,j$
with $i\geq j$,
$2\lambda_i-\lambda_j\in \Lambda$.}

\myproof{\cite[Proposition 1]{CaFaMu:arf}.}

\myproposition{
The numerical semigroup $\Lambda$ is Arf if and only if
for any $l\in\Lambda$, the set
$S(l)=\{l'-l\mid l'\in\Lambda, l'\geq l\}$ is a
numerical
semigroup.}

\myproof{Suppose $\Lambda$ is Arf.
Then $0\in S(l)$ and
if $m_1=l'-l$, $m_2=l''-l$
with $l',l''\in \Lambda$
and $l'\geq l$,
$l''\geq l$,
then $m_1+m_2=l'+l''-l-l$.
Since $\Lambda$ is Arf,
$l'+l''-l\in\Lambda$
and it is larger than or equal to $l$.
Thus, $m_1+m_2\in S(l)$.

On the other hand,
if $\Lambda$ is such that
$S(l)$
is a numerical semigroup
for any $l\in\Lambda$
then, if $\lambda_i\geq\lambda_j\geq\lambda_k$
are in $\Lambda$, we will have
$\lambda_i-\lambda_k\in S(\lambda_k)$,
$\lambda_j-\lambda_k\in S(\lambda_k)$,
$\lambda_i+\lambda_j-\lambda_k-\lambda_k\in S(\lambda_k)$
and therefore $\lambda_i+\lambda_j-\lambda_k\in\Lambda$.
}

\section{Numerical semigroups generated by an interval}
\label{section: interval numerical semigroups}

A numerical semigroup $\Lambda$
is {\it generated by an interval}
\index{semigroup!generated by an interval}
$\{i, i+1, \dots, j\}$
with $i,j\in\N_0$, $i\leq j$ if
$$\Lambda=\{n_i i+n_{i+1}(i+1)+\dots+n_j j \mid
n_i,n_{i+1},\dots,n_j\in\N_0\}.$$

A study of
semigroups
generated by intervals
was carried out
by Garc\'\i a-S\'anchez and
Rosales
in \cite{GaRo:interval}.

\myexample{
Let $q$ be a prime power.
The Hermitian curve over ${\mathbb F}_{q^2}$
is defined by the affine equation
$$x^{q+1}=y^{q}+y$$ and it has a single
rational point at infinity.
The Weierstrass semigroup at the rational point at infinity
is generated by $q$ and $q+1$
(for further details see \cite{Stichtenoth:hermite,HoLiPe:agc}).
So, it is an example
of numerical semigroup generated by an interval.
}

\mylemma{
\label{lemma: forma semigrup generat per intervals}
The semigroup $\Lambda_{\{i,\dots,j\}}$
generated by the interval $\{i,i+1,\dots,j\}$
satisfies
$$\Lambda_{\{i,\dots,j\}}=\bigcup_{k\geq 0}\{ki,ki+1,ki+2,\dots ,kj\}.$$
}

This lemma is a reformulation of
\cite[Lemma 1]{GaRo:interval}.

\myproposition{
\label{Proposition: intersection interval symmetric}
$\Lambda_{\{i,\dots,j\}}$
is symmetric
if and only if
$i\equiv 2\mod j-i.$
}

\myproof{\cite[Theorem 6]{GaRo:interval}.}

\myproposition{
The only numerical semigroups
which are generated by an interval and Arf,
are the semigroups which are equal to
$\{0\}\cup\{i\in\N_0\mid i\geq c\}$ for some non-negative integer $c$.
}

\myproof{It is a consequence of
Lemma~\ref{lemma: arf te els non-gaps separats}
and Lemma~\ref{lemma: forma semigrup generat per intervals}.
}

\myproposition{
The unique numerical semigroup
which is pseudo-symmetric and generated by an interval 
is $\{0,3,4,5,6,\dots\}$.}

\myproof{
By Lemma~\ref{lemma: forma semigrup generat per intervals},
for the non-trivial semigroup $\Lambda_{\{i,\dots,j\}}$
generated by the interval $\{i,\dots,j\}$, 
the intervals of gaps between $\lambda_1$
and the conductor satisfy that the length of each interval
is equal to the length of the previous interval minus
$j-i$.
On the other hand, the intervals of non-gaps
between $1$ and $c-1$ satisfy that the length of each interval is equal to the length of the previous interval plus $j-i$.

Now, by definition of pseudo-symmetric semigroup,
$(c-1)/2$ must be the first gap or the last gap of 
an interval of gaps.
Suppose that it is the first gap of an interval of $n$ gaps.
If it is equal to $1$ then $c=3$ and $\Lambda=\{0,3,4,5,6,\dots\}$.
Otherwise $(c-1)/2>\lambda_1$.
Then, if $\Lambda$ is pseudo-symmetric, the previous interval of non-gaps has length
$n-1$. Since $\Lambda$ is generated by an interval,
the first interval of non-gaps after 
$(c-1)/2$ must have length $n-1+j-i$ 
and since $\Lambda$ is pseudo-symmetric the interval of 
gaps before $(c-1)/2$ must have the same length. 
But since $\Lambda$ is generated by an interval,
the interval of gaps previous to $(c-1)/2$ must have length $n+j-i$.
This is a contradiction.
The same argument proves that 
$(c-1)/2$ can not be the last gap of an interval of gaps.
So, the only possibility for a pseudo-symmetric 
semigroup generated by an interval is when 
$(c-1)/2=1$, that is, when $\Lambda=\{0,3,4,5,6,\dots\}$.
}

\section{Acute numerical semigroups}
\label{section: acute numerical semigroups}

\mydefinition{We say that a numerical semigroup is {\it ordinary}
\index{semigroup!ordinary}
if it is
equal to $$\{0\}\cup\{i\in\N_0\mid i\geq c\}$$ for some non-negative integer $c$.
}

Almost all rational points on a curve of genus $g$ 
over an algebraically closed field
have
Weierstrass semigroup
of the form
$\{0\}\cup\{i\in\N_0\mid i\geq g+1\}$.
Such points are said to be {\it ordinary}.
This is why we call these numerical semigroups
ordinary \cite{Goldschmidt,FaKr,Stichtenoth:AFFaC}.
Caution must be taken when the characteristic of the ground field
is $p>0$, since there exist curves with infinitely many non-ordinary 
points \cite{StVo}.

Notice that $\N_0$ is an ordinary numerical semigroup.
It will be called the {\it trivial} numerical semigroup.
\index{semigroup!trivial}

\mydefinition{
Let $\Lambda$ be a numerical semigroup different from $\N_0$
with enumeration $\lambda$, genus $g$ and conductor $c$.
The element $\lambda_{\lambda^{-1}(c)-1}$ will be called
the {\it dominant} \index{dominant}
of the semigroup and will be denoted $d$.
For each $i\in\N_0$ let $g(i)$
be the number of gaps which are
smaller than $\lambda_i$. In particular,
$g(\lambda^{-1}(c))=g$ and
$g(\lambda^{-1}(d))=g'<g$.
If $i$ is the smallest integer for which
$g(i)=g'$ then $\lambda_i$
is called the {\it subconductor}
\index{subconductor}
of $\Lambda$ and denoted $c'$.}

\myremark{Notice that if $c'>0$, then $c'-1\not\in\Lambda$. Otherwise we would have
$g(\lambda^{-1}(c'-1))=g(\lambda^{-1}(c'))$ and $c'-1<c'$.
Notice also that all integers between $c'$ and $d$
are in $\Lambda$ because
otherwise $g(\lambda^{-1}(c'))<g'$.
}

\myremark{
\label{remark: ordinary implica d,c' son 0}
For a numerical semigroup $\Lambda$ different from $\N_0$
the following are equivalent:
\begin{description}
\item[{\bf (i)}] $\Lambda$ is ordinary,
\item[{\bf (ii)}] the dominant of $\Lambda$ is $0$,
\item[{\bf (iii)}] the subconductor of $\Lambda$ is $0$.
\end{description}
Indeed,
{\bf (i)}$\Longleftrightarrow${\bf (ii)}
and {\bf (ii)}$\Longrightarrow${\bf (iii)}
are obvious.
Now, suppose {\bf (iii)} is satisfied.
If the dominant is larger than or equal to $1$
it means that $1$ is in $\Lambda$ and so $\Lambda=\N_0$ a contradiction.
}

\mydefinition{If $\Lambda$ is a non-ordinary numerical semigroup
with enumeration $\lambda$ and
with subconductor $\lambda_i$ then
the element $\lambda_{i-1}$ will be called the {\it subdominant} \index{subdominant}
and
denoted $d'$.
}

It is well defined because of Remark~\ref{remark: ordinary implica d,c' son 0}.

\mydefinition{\label{definition:acute}
A numerical semigroup $\Lambda$
is said to be {\it acute}
\index{semigroup!acute}
if
$\Lambda$ is ordinary or if $\Lambda$ is non-ordinary and
its conductor $c$, its subconductor $c'$, its dominant $d$ and
its subdominant $d'$
satisfy
$c-d\leq c'-d'$.
}

Roughly speaking,
a numerical semigroup is acute if
the last interval of gaps before
the conductor is smaller than the previous interval of gaps.

\myexample{
\index{Hermitian!semigroup}
For the Hermitian curve over $\F_{16}$
the Weierstrass semigroup at the unique point
at infinity is
$$\{0,4,5,8,9,10\}\cup\{i\in\N_0\mid i\geq 12\}.$$
In this case
$c=12$, $d=10$, $c'=8$
and $d'=5$ and it
is easy to check that it is an acute numerical semigroup.
}

\myexample{
\index{Klein quartic!semigroup}
For the Weierstrass semigroup
at the rational point $P_0$
of the Klein quartic
in Example~\ref{example:klein}
we have
$c=5$, $d=c'=3$ and $d'=0$.
So, it is an example of a non-ordinary
acute numerical
semigroup.
}

\myproposition{
\label{proposition: exemples acute}
Let $\Lambda$ be a numerical semigroup.
\begin{enumerate}
\item If $\Lambda$ is symmetric then it is acute.
\index{semigroup!symmetric}
\item If $\Lambda$ is pseudo-symmetric then it is acute.
\item If $\Lambda$ is Arf then it is acute.
\index{semigroup!Arf}
\item If $\Lambda$ is generated by an interval then it is acute.
\index{semigroup!generated by an interval}
\end{enumerate}
}

\myproof{
If $\Lambda$ is ordinary then it is obvious. Let us suppose that $\Lambda$
is a non-ordinary semigroup
with
genus $g$, conductor $c$, subconductor $c'$, dominant $d$ and subdominant~$d'$.
\begin{enumerate}
\item Suppose that $\Lambda$ is symmetric. We know
by Proposition~\ref{proposition:symmetric-caracteritzacio}
that a numerical semigroup $\Lambda$ is symmetric if and only if
for any non-negative integer $i$, if $i$ is a gap,
then $c-1-i\in\Lambda$. If moreover it is not ordinary, then $1$ is a gap. So,
$c-2\in\Lambda$ and it is precisely the dominant.
Hence, $c-d=2$. Since $c'-1$ is a gap,
$c'-d'\geq 2=c-d$
and so $\Lambda$ is acute.
\item 
Suppose that $\Lambda$ is pseudo-symmetric.
If $1=(c-1)/2$ 
then $c=3$ and $\Lambda=\{0,3,4,5,6,\dots\}$ which is ordinary.
Else if $1\neq(c-1)/2$ then the proof 
is equivalent to the one for symmetric semigroups.
\item Suppose $\Lambda$ is Arf.
Since $d\geq c'> d'$, then
$d+c'-d'$ is in $\Lambda$
and it is strictly larger than the dominant $d$. Hence it is larger than
or equal
to $c$. So,
$d+c'-d'\geq c$
and $\Lambda$ is acute.
\item Suppose that $\Lambda$ is generated by the interval
$\{i, i+1, \dots , j\}$. Then, by
Lemma~\ref{lemma: forma semigrup generat per intervals},
there exists $k$ such that
$c=k i$, $c'=(k-1) i$, $d=(k-1) j$ and
$d'=(k-2) j$.
So, $c-d=k(i-j)+j$ while
$c'-d'=k(i-j)-i+2j$. Hence,
$\Lambda$ is acute.
\end{enumerate}
}

\begin{figure}
\centering
\setlength{\unitlength}{.5cm}
\newcommand\ambcolor[1]{#1}

{\small
\begin{picture}(16.,20.)

\put(8.,19.){\makebox(0,0){\ambcolor{\fbox{\black{Numerical semigroups}}}}}

\ambcolor{
\qbezier(8.,18.5)(8.,18.5)(8.,16.45)}

\put(8.,16.){\makebox(0,0){\ambcolor{\fbox{\black{Acute}}}}}
\ambcolor{\qbezier(8.,15.5)(8.,15.5)(8.,13.45)}

\put(8.,13.){\makebox(0,0){\ambcolor{\fbox{\black{Irreducible}}}}}
\ambcolor{
\qbezier(8.,12.5)(8.,12.5)(6.,10.5)
\qbezier(8.,12.5)(8.,12.5)(10.,10.5)}

\ambcolor{
\qbezier(8.,15.5)(8.,15.6)(2.,10.45)
\qbezier(8.,15.5)(8.,15.6)(14.,10.55)
}

\put(6.,10.){\makebox(0,0){\ambcolor{\fbox{\black{Symmetric}}}}}
\put(6.,8.8){\makebox(0,0){{\tiny\shortstack[b]{EX: Hermitian c.\\Norm-Trace c.}}}}
\put(10.,10.){\makebox(0,0){\ambcolor{\fbox{\black{Pseudo-sym}}}}}
\put(10.,8.8){\makebox(0,0){{\tiny\shortstack[b]{EX: Klein quartic.}}}}
\put(2.,10.){\makebox(0,0){\ambcolor{\fbox{\black{Arf}}}}}
\put(2.,8.8){\makebox(0,0){{\tiny\shortstack[b]{EX: Klein quartic.\\Gar-Sti  tower}}}}
\put(14.,10.){\makebox(0,0){\ambcolor{\fbox{\black{$\Lambda_{\{i,\dots,j\}}$}}}}}
\put(14.,8.8){\makebox(0,0){{\tiny\shortstack[b]{EX: Hermitian c.}}}}

\ambcolor{
\qbezier(6.,8.)(6.,8.)(4.5,6.)
\qbezier(2.,8.)(2.,8.)(8.,6.)
\qbezier(14.,8.)(14.,8.)(11.5,6.2)
\qbezier(6.,8.)(6.,8.)(11.5,6.2)
\qbezier(2.,8.)(2.,8.)(4.5,6.)
\qbezier(14.,8.)(14.,8.)(8.,6.)
\qbezier(10.,8.3)(2.,7.6)(0.7,4.9)
\qbezier(2.,8.)(2.,8.)(0.7,4.9)
\qbezier(10.,8.3)(14.,7.4)(15.5,4.5)
\qbezier(14.,8.)(14.,8.)(15.5,4.5)
}

\put(1.,4.){\makebox(0,0){\ambcolor{\fbox{\black{\shortstack[b]{Klein q.\\$\{0,3,4,\dots\}$}}}}}}
\put(4.2,5.5){\makebox(0,0){\ambcolor{\fbox{\black{Hyperelliptic}}}}}
\put(4.6,4.3){\makebox(0,0){{\tiny\shortstack[b]{Campillo, Farran,\\Munuera}}}}
\put(7.9,5.5){\makebox(0,0){\ambcolor{\fbox{\black{Ordinary}}}}}
\put(11.8,5.5){\makebox(0,0){\ambcolor{\fbox{\black{$\Lambda_{\{i,\dots,\frac{(k+1)i-2}{k}\}}$}}}}}
\put(11.6,4.3){\makebox(0,0){{\tiny\shortstack[b]{Garc\'\i a-S\'anchez,\\Rosales}}}}
\put(15.,4.){\makebox(0,0){\ambcolor{\fbox{\black{$\{0,3,4,\dots\}$}}}}}

\ambcolor{
\qbezier(4.5,3.8)(4.5,3.8)(8.,0.5)
\qbezier(8.,5.)(8.,5.)(8.,0.5)
\qbezier(11.5,3.8)(11.5,3.8)(8.,0.5)
}

\put(8.,0.){\makebox(0,0){\ambcolor{\fbox{\black{Trivial: $\N_0$}}}}}
\end{picture}
}

\caption{Diagram of semigroup classes and inclusions.}
\label{figure:inclosions}
\end{figure}
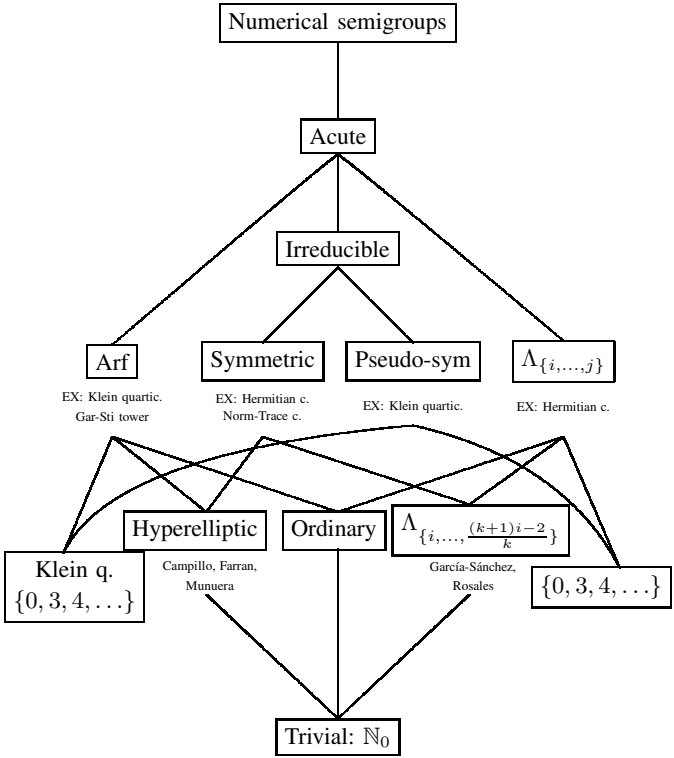

In Figure~\ref{figure:inclosions} we summarize
all the relations
we have proved
between
acute semigroups,
symmetric and pseudo-symmetric semigroups,
Arf semigroups
and semigroups generated by an interval.

\myremark{There exist numerical semigroups which are not acute.
For instance, $$\Lambda=\{0,6,8,9\}\cup\{i\in\N_0\mid i\geq 12\}.$$
In this case,
$c=12$, $d=9$, $c'=8$ and $d'=6$.

On the other hand there exist
numerical semigroups which are
acute and which are not
symmetric, pseudo-symmetric, Arf or interval-generated.
For example,
$$\Lambda=\{0,10,11\}\cup\{i\in\N_0\mid i\geq 15\}.$$
In this case,
$c=15$, $d=11$, $c'=10$ and $d'=0$.
}

\section{On the order bound on the minimum distance}
\label{section: frd acute}

In this section we will find a formula for the smallest $m$ for which 
$d_{ORD}(C_i)=\nu_{i+1}$
for all $i\geq m$, for the case of acute semigroups.
We will use the following well-known result on the values $\nu_i$.

\myproposition{
\label{proposition:nu}
Let $\Lambda$ be a numerical semigroup with genus $g$, conductor $c$ and enumeration $\lambda$.
Let $g(i)$
be the number of gaps
smaller than $\lambda_i$
and let
$$D(i)=\{l\in\Lambda^c\mid \lambda_i-l\in\Lambda^c\}.$$
Then for all $i\in\N_0$,
$$\nu_i=i-g(i)+\#D(i)+1.$$
In particular,
for all $i\geq 2c-g-1$
(or equivalently, for all $i$ such that
$\lambda_i\geq 2c-1$),
$\nu_i=i-g+1.$
}
\myproof{\cite[Theorem 3.8.]{KiPe:telescopic}.}

\myremark{
\label{remark: c'+d geq c}
Let $\Lambda$ be a non-ordinary numerical semigroup
with conductor $c$, subconductor $c'$ and
dominant $d$.
Then, $c'+d\geq c$.
Indeed,
$c'+d\in\Lambda$
and by Remark~\ref{remark: ordinary implica d,c' son 0}
it is strictly larger than $d$.
So, it must be larger than
or equal to
$c$.
}

\mytheorem{
\label{theorem: ultim punt decreixement per acute semigroups}
Let $\Lambda$ be a non-ordinary acute \index{semigroup!acute}
numerical semigroup
with enumeration $\lambda$, conductor $c$, subconductor $c'$
and
dominant $d$.
Let $$m=\min\{\lambda^{-1}(c+c'-2),
\lambda^{-1}(2d)\}.$$
Then,
\begin{enumerate}
\item $\nu_m>\nu_{m+1}$
\item $\nu_i\leq\nu_{i+1}$
for all $i>m$.
\end{enumerate}
}

\myproof{
Following the notations in 
Proposition~\ref{proposition:nu}, 
for $i\geq \lambda^{-1}(c)$,
$g(i)=g$.
Thus, for $i\geq \lambda^{-1}(c)$ we have
\begin{equation}
\label{equation: nu creixent}
\nu_i\leq\nu_{i+1}\mbox{ if and only if }
\#D(i+1)\geq \#D(i)-1.
\end{equation}

Let $l=c-d-1$.
Notice that $l$ is the number of gaps between
the conductor and the dominant.
Since $\Lambda$ is acute,
the $l$
integers before $c'$
are also gaps.
Let us call $k=\lambda^{-1}(c'+d)$.
For all $1\leq i\leq l$, both
$(c'-i)$ and $(d+i)$
are in $D(k)$ because
they are gaps and
$$(c'-i)+(d+i)=c'+d.$$
Moreover, there are no more gaps in $D(k)$ because,
if $j\leq c'-l-1$ then $c'+d-j\geq d+l+1=c$ and so
$c'+d-j\in\Lambda$.
Therefore,
$$D(k)=\{c'-i\mid 1\leq i \leq l\}\cup\{d+i\mid 1\leq i \leq l\}.$$

Now suppose that $j\geq k$.
By Remark~\ref{remark: c'+d geq c},
$\lambda_k\geq c$ and so
$\lambda_j=\lambda_k+j-k=c'+d+j-k$.
Then,
$$D(j)=A(j)\cup B(j),$$
where
{\small
\begin{eqnarray*}
A(j)&=&\casos
{\begin{array}{l}\{c'-i\mid 1\leq i \leq l-j+k\}
\\\cup\{d+i\mid j-k+1\leq i \leq l\}\end{array}}
{{\tiny if $\lambda_k\leq\lambda_{j}\leq c+c'-2$,}}
{\emptyset}{{\tiny otherwise.}}
\\
B(j)&=&\quatrecasos
{\emptyset}{{\tiny if $\lambda_k\leq\lambda_{j}\leq 2d+1$}}
{\{d+i\mid 1\leq i \leq \lambda_{j}-2d-1\}}{{\tiny if $2d+2\leq\lambda_{j}\leq c+d$,}}
{\{d+i\mid \lambda_{j}-d-c+1\leq i \leq l\}}
{{\tiny if $c+d\leq\lambda_{j}\leq 2c-2$,}}
{\emptyset}{{\tiny if $\lambda_{j}\geq 2c-1$.}}
\end{eqnarray*}}
Notice that $A(j)\cap B(j)=\emptyset$ and hence
$$\#D(j)=\#A(j)+\#B(j).$$
We have
\begin{eqnarray*}
\#A(j)&=&\casos
{2(l-j+k)}{if $\lambda_k\leq \lambda_{j}\leq c+c'-2$,}
{0}{otherwise.}\\
\#B(j)&=&\quatrecasos
{0}{if $\lambda_k\leq \lambda_{j}\leq 2d+1$,}
{\lambda_{j}-2d-1}{if $2d+2\leq\lambda_{j}\leq c+d$,}
{2c-1-\lambda_j}{if $c+d\leq\lambda_{j}\leq 2c-2$,}
{0}{if $\lambda_{j}\geq 2c-1$.}
\end{eqnarray*}
So,
\begin{eqnarray*}
\#A(j+1)&=&\casos{\#A(j)-2}{if $\lambda_k\leq
\lambda_{j}\leq c+c'-2$,}{\#A(j)}{otherwise.}
\\
\#B(j+1)&=&\quatrecasos
{\#B(j)}{if $\lambda_k\leq\lambda_{j}\leq 2d$}
{\#B(j)+1}{if $2d+1\leq\lambda_{j}\leq c+d-1$}
{\#B(j)-1}{if $c+d\leq\lambda_{j}\leq 2c-2$}
{\#B(j)}{if $\lambda_{j}\geq 2c-1$}
\end{eqnarray*}
Notice that $c+c'-2< c+d$.
Thus, for $\lambda_j \geq c+d$,
$$\#D(j+1)=\casos{\#D(j)-1}{if $c+d\leq\lambda_{j}\leq 2c-2$,}
{\#D(j)}{if $\lambda_{j}\geq 2c-1$.}$$
Hence, by (\ref{equation: nu creixent}),
$\nu_i\leq \nu_{i+1}$ for all $i\geq \lambda^{-1}(c+d)$
because $\lambda^{-1}(c+d)\geq\lambda^{-1}(c)$.
Now, let us analyze what happens if
$\lambda_{j}< c+d$.

If $c+c'-2\leq 2d$ then
$$\#D(j+1)=\trescasos
{\#D(j)-2}{if $\lambda_k\leq \lambda_{j}\leq c+c'-2$,}
{\#D(j)}{if $c+c'-1\leq\lambda_{j}\leq 2d$,}
{\#D(j)+1}{if $2d+1\leq\lambda_{j}\leq c+d-1$.}$$
and if
$2d +1\leq c+c'-2$ then
$$\#D(j+1)=\trescasos
{\#D(j)-2}{if $\lambda_k\leq \lambda_{j}\leq 2d$,}
{\#D(j)-1}{if $2d+1\leq\lambda_{j}\leq c+c'-2$,}
{\#D(j)+1}{if $c+c'-1\leq\lambda_{j}\leq c+d-1$.}$$

So, by (\ref{equation: nu creixent})
and since both $c+c'-2$ and $2d$ are larger than or equal to $c$,
the result follows.
}

\mycorollary{
\label{corollary:dFR-acute}
Let $\Lambda$ be a non-ordinary acute numerical semigroup
with enumeration $\lambda$, conductor $c$ and subconductor $c'$.
Let $$m=\min\{\lambda^{-1}(c+c'-2),
\lambda^{-1}(2d)\}.$$
Then, $m$ is the smallest integer
for which $$d_{ORD}(C_i)=\nu_{i+1}$$
for all $i\geq m$.}

\myexample{
Recall the
Weierstrass semigroup at the point $P_0$
on the Klein quartic that we presented
in Example~\ref{example:klein}.
Its conductor is $5$,
its dominant is $3$
and its subconductor is $3$.
In Table~\ref{table:klein} we have,
for each integer from $0$ to
$\lambda^{-1}(2c-2)$,
the values $\lambda_i$, $\nu_i$ and $d_{ORD}(C_i)$.
Recall that,
as mentioned in the introduction,
$\nu_{i+1}\leq \nu_{i+2}$ and
$d_{ORD}(C_i)=\nu_{i+1}$ for all $i\geq \lambda^{-1}(2c-1)-1$.

For this example,
$\lambda^{-1}(c+c'-2)=\lambda^{-1}(2d)=3$
and so,
$m=\min\{\lambda^{-1}(c+c'-2),
\lambda^{-1}(2d)\}=3$.
We can check that, as
stated in
Theorem~\ref{theorem: ultim punt decreixement per acute semigroups},
$\nu_3>\nu_4$
and $\nu_i\leq \nu_{i+1}$ for all $i>3$.
Moreover,
as stated in Corollary~\ref{corollary:dFR-acute},
$d_{ORD}(C_i)=\nu_{i+1}$
for all $i\geq 3$
while $d_{ORD}(C_2)\neq\nu_3$.
}

\begin{table}
\caption{Klein quartic}
\label{table:klein}
\centering
\begin{tabular}{|c|c|c|c|}
\hline
$i$ & $\lambda_i$ & $\nu_i$ & $d_{ORD}(C_i)$ \\ \hline
$0$ & $0$ & $1$ & $2$ \\
$1$ & $3$ & $2$ & $2$ \\
$2$ & $5$ & $2$ & $2$ \\
$3$ & $6$ & $3$ & $2$ \\
$4$ & $7$ & $2$ & $4$ \\
$5$ & $8$ & $4$ & $4$ \\
\hline \end{tabular}

\end{table}

\myproposition{
\label{proposition: qui es el minim dels dos}
Let $\Lambda$ be a non-ordinary numerical semigroup
with conductor $c$, subconductor $c'$ and dominant~$d$.
\begin{enumerate}
\item If $\Lambda$ is symmetric then $\min\{c+c'-2,2d\}=c+c'-2=2c-2-\lambda_1$,
\item If $\Lambda$ is pseudo-symmetric then $\min\{c+c'-2,2d\}=c+c'-2$,
\item If $\Lambda$ is Arf then $\min\{c+c'-2,2d\}=2d$,
\item If $\Lambda$ is generated by an interval then $\min\{c+c'-2,2d\}=c+c'-2$.
\end{enumerate}
}

\myproof{\mbox{}
\begin{enumerate}
\item We already saw in the proof
of Proposition~\ref{proposition: exemples acute}
that if
$\Lambda$ is symmetric then
$d=c-2$. So, $c+c'-2=d+c'\leq 2d$ because
$c'\leq d$.
Moreover, by Proposition~\ref{proposition:symmetric-implicacio},
any non-negative integer $i$ is a gap if and only if $c-1-i\in\Lambda$.
This implies that
$c'-1=c-1-\lambda_1$ and so
$c'=c-\lambda_1$. Therefore,
$c+c'-2=2c-2-\lambda_1$.
\item If $\Lambda$ is pseudo-symmetric and non-ordinary then
$d=c-2$ because $1$ is a gap different from $(c-1)/2$. 
So, $c+c'-2=d+c'\leq 2d$.
\item If $\Lambda$ is Arf then
$c'=d$. Indeed, if $c'<d$ then
$d-1\in\Lambda$ and, since $\Lambda$ is Arf,
$d+1=2d-(d-1)\in\Lambda$, a contradiction.
Since $d\leq c-2$, we have $2d\leq c+c'-2$.
\item Suppose $\Lambda$ is generated by the interval
$\{i,i+1,\dots,j\}$.
By Lemma~\ref{lemma: forma semigrup generat per intervals},
there exists $k$ such that
$c=k i$ and $d=(k-1) j$.
We have that $c-d\leq j-i$, because otherwise
$(k+1)i-k j=c-d-(j-i)>0$, and hence
$k j+1$ would be a gap greater than $c$.
On the other hand
$d-c'\geq j-i$,
and hence
$2d-(c+c'-2)=d-c+d-c'+2\geq i-j+j-i+2\geq 2$.
\end{enumerate}
}

\myexample{
Consider the Hermitian curve
\index{Hermitian!semigroup}
over $\F_{16}$.
Its numerical semigroup is generated by $4$ and $5$.
So, this is a symmetric numerical semigroup
because it is generated by two coprime integers,
and it is
also a semigroup generated by the interval
$\{4,5\}$.

In Table~\ref{table:hermite} we include,
for each integer from $0$ to $16$,
the values $\lambda_i$, $\nu_i$ and $d_{ORD}(C_i)$.
Notice that in this case the conductor is $12$,
the dominant is $10$ and the subconductor is $8$.
We do not give the values
in the table for
$i> \lambda^{-1}(2c-1)-1=16$
because
$d_{ORD}(C_i)=\nu_{i+1}$ for all $i\geq \lambda^{-1}(2c-1)-1$.
We can check that, as
follows from Theorem~\ref{theorem: ultim punt decreixement per acute semigroups}
and Proposition~\ref{proposition: qui es el minim dels dos},
$\lambda^{-1}(c+c'-2)=12$ is the largest integer $m$
with $\nu_m>\nu_{m+1}$ and so
the smallest integer for which
$d_{ORD}(C_i)=\nu_{i+1}$
for all $i\geq m$.
Notice also that, as
pointed out in Proposition~\ref{proposition: qui es el minim dels dos},
$c+c'-2=2c-2-\lambda_1$.

Furthermore, in this example
there are $64$ rational points
on the curve different from $P_\infty$ 
and the map $\varphi$
evaluating the functions of $A$ at these $64$ points
satisfies
that the words $\varphi(f_0),\dots,\varphi(f_{57})$
are linearly independent whereas $\varphi(f_{58})$
is linearly dependent to the previous ones.
So, $d_{ORD}^\varphi(C_i)=d_{ORD}(C_i)$
for all $i\leq 56$.
}

\begin{table}
\caption{Hermitian curve}
\label{table:hermite}
\centering
\begin{tabular}{|c|c|c|c|}
\hline
$i$ & $\lambda_i$ & $\nu_i$ & $d_{ORD}(C_i)$ \\ \hline
$0$ & $0$ & $1$ & $2$ \\
$1$ & $4$ & $2$ & $2$ \\
$2$ & $5$ & $2$ & $3$ \\
$3$ & $8$ & $3$ & $3$ \\
$4$ & $9$ & $4$ & $3$ \\
$5$ & $10$ & $3$ & $4$ \\
$6$ & $12$ & $4$ & $4$ \\
$7$ & $13$ & $6$ & $4$ \\
$8$ & $14$ & $6$ & $4$ \\
$9$ & $15$ & $4$ & $5$ \\
$10$ & $16$ & $5$ & $8$ \\
$11$ & $17$ & $8$ & $8$ \\
$12$ & $18$ & $9$ & $8$ \\
$13$ & $19$ & $8$ & $9$ \\
$14$ & $20$ & $9$ & $10$ \\
$15$ & $21$ & $10$ & $12$ \\
$16$ & $22$ & $12$ & $12$ \\
\hline \end{tabular}

\end{table}

\myexample{
Let us consider now the
semigroup of the fifth code associated to the second tower of
Garcia and Stichtenoth over $\F_{4}$.
\index{Garcia-Stichtenoth towers!semigroup}
As noticed
in Example~\ref{example:2nd GS es Arf},
this is an Arf numerical semigroup.
We set in Table~\ref{table:garciastichtenoth}
the values $\lambda_i$, $\nu_i$ and $d_{ORD}(C_i)$
for each integer from $0$ to $25$.
In this case the conductor is $24$,
the dominant is $20$ and the subconductor is $20$.
As before, we do not give the values for
$i> \lambda^{-1}(2c-1)-1=25$.
We can check that, as
follows from Theorem~\ref{theorem: ultim punt decreixement per acute semigroups}
and Proposition~\ref{proposition: qui es el minim dels dos},
$\lambda^{-1}(2d)=19$ is the largest integer $m$
with $\nu_m>\nu_{m+1}$ and so,
the smallest integer for which
$d_{ORD}(C_i)=\nu_{i+1}$
for all $i\geq m$.
}

\begin{table}
\caption{Garcia-Stichtenoth tower}
\label{table:garciastichtenoth}
\centering
\begin{tabular}{|c|c|c|c|}
\hline
$i$ & $\lambda_i$ & $\nu_i$ & $d_{ORD}(C_i)$ \\ \hline
$0$ & $0$ & $1$ & $2$ \\
$1$ & $16$ & $2$ & $2$ \\
$2$ & $20$ & $2$ & $2$ \\
$3$ & $24$ & $2$ & $2$ \\
$4$ & $25$ & $2$ & $2$ \\
$5$ & $26$ & $2$ & $2$ \\
$6$ & $27$ & $2$ & $2$ \\
$7$ & $28$ & $2$ & $2$ \\
$8$ & $29$ & $2$ & $2$ \\
$9$ & $30$ & $2$ & $2$ \\
$10$ & $31$ & $2$ & $2$ \\
$11$ & $32$ & $3$ & $2$ \\
$12$ & $33$ & $2$ & $2$ \\
$13$ & $34$ & $2$ & $2$ \\
$14$ & $35$ & $2$ & $2$ \\
$15$ & $36$ & $4$ & $2$ \\
$16$ & $37$ & $2$ & $2$ \\
$17$ & $38$ & $2$ & $2$ \\
$18$ & $39$ & $2$ & $4$ \\
$19$ & $40$ & $5$ & $4$ \\
$20$ & $41$ & $4$ & $4$ \\
$21$ & $42$ & $4$ & $4$ \\
$22$ & $43$ & $4$ & $6$ \\
$23$ & $44$ & $6$ & $6$ \\
$24$ & $45$ & $6$ & $6$ \\
$25$ & $46$ & $6$ & $6$ \\
\hline \end{tabular}

\end{table}

\section{On the improvement of the codes $\widetilde{C}(\delta)$}
\label{section: nu no decreasing implica ordinary}

\myproposition{
\label{proposition: nui ordinary}
If $\Lambda$ is an ordinary numerical semigroup with enumeration $\lambda$
then
$$\nu_i=\trescasos{1}{if $i=0$,}{2}{if $1\leq i\leq \lambda_1$,}
{i-\lambda_1+2}{if $i>\lambda_1$.}$$
\bigskip
}

\myproof{It is obvious that $\nu_0=1$
and that $\nu_i=2$ whenever
$0<\lambda_i<2\lambda_1$.
So, since $2\lambda_1=\lambda_{\lambda_1+1}$,
we have that $\nu_i=2$ for all $1\leq i\leq \lambda_1$.
Finally, if $\lambda_i\geq 2\lambda_1$ then
all non-gaps up to $\lambda_i-\lambda_1$
are in $N_i$ as well as $\lambda_i$, and
none of the remaining non-gaps are in $N_i$.
Now, if the genus of $\Lambda$ is $g$, then
$\nu_i=\lambda_i-\lambda_1+2-g$
and
$\lambda_i=i+g$.
So,
$\nu_i=i-\lambda_1+2$.
}

As a consequence of
Proposition~\ref{proposition: nui ordinary},
the sequence $(\nu_i)$ is non-decreasing if $\Lambda$
is an ordinary numerical semigroup.
We will see in this section that
ordinary numerical semigroups are in fact the only semigroups for
which $(\nu_i)$ is non-decreasing.

\mylemma{
\label{lemma: nu non-decreasing implica Arf}
Suppose that for the semigroup $\Lambda$ the sequence
$(\nu_i)$ is non-de\-crea\-sing. Then $\Lambda$ is Arf.}

\myproof{
Let $\lambda$ be the enumeration of $\Lambda$.
Let us see by induction that, for any non-negative integer $i$,
\begin{description}
\item[{\bf (i)}]
$N_{\lambda^{-1}(2\lambda_i)}=
  \{j\in\N_0\mid j\leq i\}\sqcup\{\lambda^{-1}(2\lambda_i-\lambda_j)\mid 0\leq j<i\}$,
where $\sqcup$ means the union
of disjoint sets.
\item[{\bf (ii)}]
$N_{\lambda^{-1}(\lambda_i+\lambda_{i+1})}=
\{j\in\N_0\mid j\leq i\}\sqcup\{\lambda^{-1}(\lambda_i+\lambda_{i+1}-\lambda_j)\mid 0\leq j\leq i\}$.
\end{description}
Notice that if {\bf (i)}
is satisfied for all $i$,
then
$\{j\in\N_0\mid j\leq i\}\subseteq N_{\lambda^{-1}(2\lambda_i)}$
for all $i$, and hence
by
Proposition~\ref{proposition: charactarf 2j-i}
$\Lambda$ is Arf.

It is obvious
that both {\bf (i)} and {\bf (ii)}
are satisfied
for the case $i=0$.

Suppose $i>0$.
By the induction hypothesis,
$\nu_{\lambda^{-1}(\lambda_{i-1}+\lambda_i)}=2i$.
Now, since $(\nu_i)$ is not decreasing
and $2\lambda_i>\lambda_{i-1}+\lambda_i$,
we have
$\nu_{\lambda^{-1}(2\lambda_i)}\geq 2i$.
On the other hand,
if $j,k\in\N_0$ are such that $j\leq k$ and $\lambda_j+\lambda_k=2\lambda_i$
then $\lambda_j\leq\lambda_i$ and $\lambda_k\geq \lambda_i$.
So,
$\lambda(N_{\lambda^{-1}(2\lambda_i)})\subseteq
\{\lambda_j\mid 0\leq j\leq i\}\sqcup\{2\lambda_i-\lambda_j\mid 0\leq j<i\}$
and hence
$\nu_{\lambda^{-1}(2\lambda_i)}\geq 2i$ if and
only if
$N_{\lambda^{-1}(2\lambda_i)}=
\{j\in\N_0\mid j\leq i\}\sqcup\{\lambda^{-1}(2\lambda_i-\lambda_j)\mid 0\leq j<i\}$.
This proves {\bf (i)}.

Finally, {\bf (i)} implies
$\nu_{\lambda^{-1}(2\lambda_i)}=2i+1$
and {\bf (ii)} follows by an analogous argumentation.
}

\mytheorem{
\label{theorem: non-decreasing nu implica ordinary}
The only numerical semigroups for which
$(\nu_i)$
is non-de\-crea\-sing are ordinary numerical semigroups.
\index{semigroup!ordinary}}

\myproof{
It is a consequence of Lemma~\ref{lemma: nu non-decreasing implica Arf},
Proposition~\ref{proposition: exemples acute},
Theorem~\ref{theorem: ultim punt decreixement per acute semigroups}
and Proposition~\ref{proposition: nui ordinary}.
}

\mycorollary{\label{corollary: nu increasing implica semigrup trivial}
The only numerical semigroup for which
$(\nu_i)$
is strictly increasing is the trivial numerical semigroup.
}

\myproof{It is a consequence of
Theorem~\ref{theorem: non-decreasing nu implica ordinary}
and Proposition~\ref{proposition: nui ordinary}.
}

As a consequence of 
Theorem~\ref{theorem: non-decreasing nu implica ordinary}
we can show that the only numerical semigroups
for which the related evaluation codes
$C(\delta)$
are not improved 
by the codes
$\widetilde{C}(\delta)$,
at least for one value of the
designed minimum distance, are
ordinary semigroups.

\mycorollary{
Given a numerical semigroup $\Lambda$ define
$m(\delta)=\max\{i\in\N_0\mid \nu_i<\delta\}$.
There exists at least one value of $\delta$
for which $\{i\in\N_0\mid\nu_i<\delta\}\subsetneq\{i\in\N_0\mid i\leq m(\delta)\}$
if and only if $\Lambda$ is non-ordinary.
}

\section{The sequence $(\nu_i)$ determines the semigroup}
\label{section: nu determina lambda}

\mytheorem{\label{theorem: nu determina lambda}
Suppose that $(\nu_i)$ corresponds to the numerical
semigroup $\Lambda$. Then there is no other numerical semigroup
with the same sequence $(\nu_i)$.}

\myproof{If $\Lambda=\N_0$ then $(\nu_i)$ is strictly increasing and,
by Corollary~\ref{corollary: nu increasing implica semigrup trivial}, there
is no other semigroup with the same sequence $(\nu_i)$.

Suppose that $\Lambda$ is not trivial.
Then we can determine the genus and the conductor
from the sequence $(\nu_i)$.
Indeed, let $k=2c-g-2$. 
In the following we will show how to determine
$k$ without the knowledge of $c$ and $g$.
Notice that
$c\geq 2$ and so
$2c-2\geq c$.
This implies $k=\lambda^{-1}(2c-2)$ and $g(k)=g$.
By Proposition~\ref{proposition:nu},
$\nu_{k}=k-g+\#D(k)+1$.
But $D(k)=\{c-1\}$.
So, $\nu_k=k-g+2$.
By Proposition~\ref{proposition:nu} again,
$\nu_i=i-g+1$ for all $i>k$ and so
we have
$$k=\max\{i\mid \nu_i=\nu_{i+1}\}.$$
We can determine the genus as
$$g=k+2-\nu_{k}$$
and the conductor as
$$c=\frac{k+g+2}{2}.$$
Now we know that $\{0\}\in\Lambda$
and $\{i\in\N_0\mid i\geq c\}\subseteq\Lambda$
and, furthermore,
$\{1,c-1\}\subseteq\Lambda^{c}$.
It remains to determine 
for all $i\in\{2,\dots,c-2\}$
whether
$i\in\Lambda$. 
Let us assume
$i\in\{2,\dots,c-2\}$.

On one hand, $c-1+i-g>c-g$ and so
$\lambda_{c-1+i-g}>c$.
This means that
$g(c-1+i-g)=g$ and hence
\begin{equation}
\label{equacio: demo nu determina lambda 1}
\nu_{c-1+i-g}=c-1+i-g-g+\#D(c-1+i-g)+1.
\end{equation}

On the other hand, if we define
$\tilde{D}(i)$ to be
$$\tilde{D}(i)=\{l\in\Lambda^{c}\mid c-1+i-l\in\Lambda^{c},i<l<c-1\}$$
then
\begin{equation}
\label{equacio: demo nu determina lambda 2}
D(c-1+i-g)=\casos{\tilde{D}(i)\cup\{c-1,i\}}{if $i\in\Lambda^c$}{\tilde{D}(i)}{otherwise.}
\end{equation}
So, from (\ref{equacio: demo nu determina lambda 1}) and
(\ref{equacio: demo nu determina lambda 2}),
\begin{center}
$i$ is a non-gap $\Longleftrightarrow
\nu_{c-1+i-g}=c+i-2g+\#\tilde{D}(i).$
\end{center}
This gives an inductive procedure to decide
whether $i$ belongs to $\Lambda$
decreasingly from $i=c-2$ to $i=2$.
}

\myremark{
From the proof of
Theorem~\ref{theorem: nu determina lambda}
we see that a semigroup can be determined by
$k=\max\{i\mid \nu_i=\nu_{i+1}\}$ and
the values $\nu_i$ for $i\in\{c-g+1,\dots,2c-g-3\}$.
}

\section*{Acknowledgments}

The author would like to thank 
Michael E. O'Sullivan, 
Ruud Pellikaan
and Pedro A. Garc\'\i a-S\'anchez
for many helpful discussions.
She would like to thank also the referees for
their careful reading and for many intresting 
remarks.

\bibliographystyle{IEEE}

\begin{thebibliography}{10}

\bibitem{BaDoFo}
Valentina Barucci, David~E. Dobbs, and Marco Fontana.
\newblock Maximality properties in numerical semigroups and applications to
  one-dimensional analytically irreducible local domains.
\newblock {\em Mem. Amer. Math. Soc.}, 125(598):x+78, 1997.

\bibitem{Bras:AAECC}
Maria Bras-Amor\'{o}s.
\newblock Improvements to evaluation codes and new characterizations of {A}rf
  semigroups.
\newblock In {\em Applied algebra, algebraic algorithms and error-correcting
  codes (Toulouse, 2003)}, Lecture Notes in Comput. Sci. Springer, 2003.

\bibitem{CaFa:semigroup_singular_plane_models}
A.~Campillo and J.~I. Farr{\'a}n.
\newblock Computing {W}eierstrass semigroups and the {F}eng-{R}ao distance from
  singular plane models.
\newblock {\em Finite Fields Appl.}, 6(1):71--92, 2000.

\bibitem{CaFaMu:arf}
Antonio Campillo, Jos{\'e}~Ignacio Farr{\'a}n, and Carlos Munuera.
\newblock On the parameters of algebraic-geometry codes related to {A}rf
  semigroups.
\newblock {\em IEEE Trans. Inform. Theory}, 46(7):2634--2638, 2000.

\bibitem{FaKr}
H.~M. Farkas and I.~Kra.
\newblock {\em Riemann surfaces}, volume~71 of {\em Graduate Texts in
  Mathematics}.
\newblock Springer-Verlag, New York, second edition, 1992.

\bibitem{FeRa:dFR}
Gui~Liang Feng and T.~R.~N. Rao.
\newblock A simple approach for construction of algebraic-geometric codes from
  affine plane curves.
\newblock {\em IEEE Trans. Inform. Theory}, 40(4):1003--1012, 1994.

\bibitem{FeRa:improved}
Gui-Liang Feng and T.~R.~N. Rao.
\newblock Improved geometric {G}oppa codes. {I}. {B}asic theory.
\newblock {\em IEEE Trans. Inform. Theory}, 41(6, part 1):1678--1693, 1995.
\newblock Special issue on algebraic geometry codes.

\bibitem{GaSt:tff}
Arnaldo Garcia and Henning Stichtenoth.
\newblock On the asymptotic behaviour of some towers of function fields over
  finite fields.
\newblock {\em J. Number Theory}, 61(2):248--273, 1996.

\bibitem{GaRo:interval}
P.~A. Garc{\'{\i}}a-S{\'a}nchez and J.~C. Rosales.
\newblock Numerical semigroups generated by intervals.
\newblock {\em Pacific J. Math.}, 191(1):75--83, 1999.

\bibitem{Geil:Norm-trace-codes}
Olav Geil.
\newblock On codes from norm-trace curves.
\newblock To appear in Finite Fields and Their Applications, 2002.

\bibitem{Goldschmidt}
David~M. Goldschmidt.
\newblock {\em Algebraic functions and projective curves}, volume 215 of {\em
  Graduate Texts in Mathematics}.
\newblock Springer-Verlag, New York, 2003.

\bibitem{HoLiPe:agc}
Tom H\o{holdt}, Jacobus~H. van Lint, and Ruud Pellikaan.
\newblock {\em Algebraic Geometry codes}, pages 871--961.
\newblock North-Holland, Amsterdam, 1998.

\bibitem{KiPe:telescopic}
Christoph Kirfel and Ruud Pellikaan.
\newblock The minimum distance of codes in an array coming from telescopic
  semigroups.
\newblock {\em IEEE Trans. Inform. Theory}, 41(6, part 1):1720--1732, 1995.
\newblock Special issue on algebraic geometry codes.

\bibitem{PeStTo}
Ruud Pellikaan, Henning Stichtenoth, and Fernando Torres.
\newblock Weierstrass semigroups in an asymptotically good tower of function
  fields.
\newblock {\em Finite Fields Appl.}, 4(4):381--392, 1998.

\bibitem{PeTo:redundancy_improved_codes}
Ruud Pellikaan and Fernando Torres.
\newblock On {W}eierstrass semigroups and the redundancy of improved geometric
  {G}oppa codes.
\newblock {\em IEEE Trans. Inform. Theory}, 45(7):2512--2519, 1999.

\bibitem{Pretzel}
Oliver Pretzel.
\newblock {\em Codes and algebraic curves}.
\newblock The Clarendon Press Oxford University Press, New York, 1998.

\bibitem{RoBr:irreducible}
J.~C. Rosales and M.~B. Branco.
\newblock Irreducible numerical semigroups.
\newblock {\em Pacific J. Math.}, 209(1):131--143, 2003.

\bibitem{RoGaGaBr}
J.~C. Rosales, P.~A. Garc\'{\i}a-S\'{a}nchez, J.~I. Garc\'{\i}a-Garc\'{\i}a,
  and M.~B. Branco.
\newblock Arf numerical semigroups.
\newblock Submitted.

\bibitem{Stichtenoth:hermite}
Henning Stichtenoth.
\newblock A note on {H}ermitian codes over {${\rm GF}(q\sp 2)$}.
\newblock {\em IEEE Trans. Inform. Theory}, 34(5, part 2):1345--1348, 1988.
\newblock Coding techniques and coding theory.

\bibitem{Stichtenoth:AFFaC}
Henning Stichtenoth.
\newblock {\em Algebraic function fields and codes}.
\newblock Springer-Verlag, Berlin, 1993.

\bibitem{StVo}
Karl-Otto St{\"o}hr and Jos{\'e}~Felipe Voloch.
\newblock Weierstrass points and curves over finite fields.
\newblock {\em Proc. London Math. Soc. (3)}, 52(1):1--19, 1986.

\end{thebibliography}

\def\cprime{$'$} \def\lfhook#1{\setbox0=\hbox{#1}{\ooalign{\hidewidth
  \lower1.5ex\hbox{'}\hidewidth\crcr\unhbox0}}}

\end{document}